\documentclass{aa}
\usepackage{graphicx}
\usepackage{placeins}
\usepackage[varg]{txfonts}
\usepackage{natbib,twoopt}
\usepackage[breaklinks=true]{hyperref} %% to avoid \citeads line fills
\bibpunct{(}{)}{;}{a}{}{,}             %% natbib format for A&A and ApJ
\makeatletter
  \newcommandtwoopt{\citeads}[3][][]{\href{http://adsabs.harvard.edu/abs/#3}%
    {\def\hyper@linkstart##1##2{}%
     \let\hyper@linkend\@empty\citealp[#1][#2]{#3}}}
  \newcommandtwoopt{\citepads}[3][][]{\href{http://adsabs.harvard.edu/abs/#3}%
    {\def\hyper@linkstart##1##2{}%
     \let\hyper@linkend\@empty\citep[#1][#2]{#3}}}
  \newcommandtwoopt{\citetads}[3][][]{\href{http://adsabs.harvard.edu/abs/#3}%
    {\def\hyper@linkstart##1##2{}%
     \let\hyper@linkend\@empty\citet[#1][#2]{#3}}}
  \newcommandtwoopt{\citeyearads}[3][][]%
    {\href{http://adsabs.harvard.edu/abs/#3}
    {\def\hyper@linkstart##1##2{}%
     \let\hyper@linkend\@empty\citeyear[#1][#2]{#3}}}
\makeatother
\usepackage{makecell}
\usepackage{siunitx}
\ifdefined\qty\else
  \ifdefined\NewCommandCopy
    \NewCommandCopy\qty\SI
  \else
    \NewDocumentCommand\qty{O{}mm}{\SI[#1]{#2}{#3}}
  \fi
\fi
\ifdefined\unit\else
  \ifdefined\NewCommandCopy
    \NewCommandCopy\unit\si
  \else
    \NewDocumentCommand\unit{O{}m}{\si[#1]{#2}}
  \fi
\fi

\DeclareSIUnit\angstrom{\text {Å}}
\DeclareSIUnit \minutes {minutes}
\DeclareSIUnit \years {years}
\DeclareSIUnit \parsec {pc}
\DeclareSIUnit \megasec {Msec}
\DeclareSIUnit \mujansky {\si{\micro Jy}}
\DeclareSIUnit \sqarcsec {$\text{arcsec}^2$}
\begin{document}

\title{The time-variable ultraviolet sky: Active Galactic Nuclei, Stars and White Dwarfs}

\author{R. Bühler
      \inst{1}
      \and
      J. Schliwinski
      \inst{2}
      }

\institute{Deutsches Elektronen-Synchrotron DESY,
          Platanenallee 6, 15735 Zeuthen, Germany\\
          \email{rolf.buehler@desy.de}
     \and
         Institut für Physik, Humboldt-Universität zu Berlin,
         Newtonstrasse 15, 12489 Berlin, Germany\\
         \email{julian.schliwinksi@desy.de}
         }

\date{Received date / Accepted date }
 
\abstract{Here, we present the 1UVA catalog of time variable ultraviolet (UV) sources. We describe a new analysis pipeline, the VAriable Source Clustering Analysis (VASCA).  We apply the pipeline to 10 years of data from the GALaxy Evolution eXplorer (GALEX) satellite. We analyse a sky area of $\qty{302}{\deg}^2$, resulting in the detection of 4202 time-variable UV sources. We cross correlate these sources with multi-frequency data from the \textit{Gaia} satellite and the SIMBAD database, finding an association for 3655 sources. The source sample is dominated by Active Galactic Nuclei (\qty{\approx 73}{\percent}) and stars (\qty{\approx 24}{\percent}). We look at UV and multi-frequency properties of these sources, focusing on the stellar population. We find UV variability for four White Dwarfs. One of them, WD~J004917.14-252556.81, has recently been found to be the most massive pulsating White Dwarf. Its Spectral Energy Distribution shows no sign of a stellar companion. The observed flux variability is unexpected and difficult to explain.}

% maximum of 6 key words
% allowed keys: https://www.aanda.org/for-authors/latex-issues/information-files#pop

\maketitle
%
%-------------------------------------------------------------------

\section{Introduction} \label{sec:intro}

The study of the time variable sky has historically been a key area in astronomy. The characterization of planet movements and the light emitted by distant supernovae, for example, have fundamentally shaped our understanding of the Universe. More recently, the time variability of stars due to the occultation of orbiting planets, has led to the discovery of thousands of extra-solar planets \citepads{2021ARA&A..59..291Z}. Over the past decades, several new classes of variable sources have been found, such as Fast Radio Bursts \citepads{2019ARA&A..57..417C}, Tidal Disruption Events \citepads{2021ARA&A..59...21G}, Kilonovae \citepads{2017ApJ...848L..12A} or Pulsar Wind Nebulae \citepads{2014RPPh...77f6901B}.

Wide field surveys have characterized the time variability of the sky from radio to gamma-ray frequencies (\citeads{2011ApJ...742...49T}; \citeads{2019PASP..131a8002B}; \citeads{2014ApJ...786...20L}; \citeads{2017ApJ...846...34A}). At all wavebands, AGN and/or variable stars make up the bulk of variable sources. The variability observed from stars typically has a thermal photon spectrum, indicating time-variable heating and/or cooling of the star or its environment. For AGN, the variability often shows non-thermal spectra and is therefore linked to acceleration and cooling of cosmic rays. However, exceptions to both of these generalizations exist.

In this article, we study the variability of the UV sky, using data from the GALaxy Evolution eXplorer (GALEX). This satellite scanned \qty{\approx 70}{\percent} of the sky from 2003 to 2013. To date, its data is still the one with the best UV-coverage in time over a wide field. GALEX took data in two filters, in the NUV ($\lambda_{eff} = \qty{2316}{\angstrom}$) and FUV ($\lambda_{eff} = \qty{1539}{\angstrom}$) bands. The FUV sensor failed in 2009, which is why only NUV data is available after this. A summary of the GALEX instrument performance can be found in \citetads{2007ApJS..173..682M} and a description of its different surveys in \citetads{2014AdSpR..53..900B}. 

The most detailed systematic characterization of time-variable sources in the GALEX data was done in the Time Domain Survey (TDS, \citeads{2013ApJ...766...60G}). The TDS covered an area of \qty{40}{\square\deg} finding 1078 UV-variable sources. More recently, a project has been started to create a legacy catalog of GALEX sources using all available GALEX data\footnote{\url{https://www.millionconcepts.com/documents/glcat_adap_trimmed.pdf}}. This catalog shall also include time variability information. As a first result, a catalog of 1426 sources which vary on small time scales of \qty{\lessapprox 1500}{\sec} was derived \citepads{2023ApJS..268...41M}. Future missions, as ULTRASAT \citepads{2023arXiv230414482S}, the CSST \citepads{2018cosp...42E3821Z} and the proposed UVEX \citepads{2021arXiv211115608K} are expected to improve the sensitivity of such studies in the coming decade.

In this work, we created the 1UVA catalog of variable sources from the GALEX data, extending the sky coverage of the TDS by a factor $\gtrapprox 7$. For this, we implemented a new analysis pipeline, the VAriable Source Cluster Analysis (VASCA). The article is structured as follows: In section \ref{sec:vasca}, we describe the VASCA pipeline, the GALEX dataset and the source association procedures. In section \ref{sec:catalog}, we present the obtained results and discuss the source classes, focusing on UV variable stars and White Dwarfs (WDs). Finally, we summarize our findings in section \ref{sec:summary}. 

The VASCA code is publicly available on GitHub\footnote{\url{https://github.com/rbuehler/vasca}} and the data products of the 1UVA catalog will be made available at the Strasbourg astronomical Data Center (CDS). Throughout this paper we will report spectral flux density in micro Jansky, AB magnitudes will also be given in parallel for comparison with other works.

%
%-------------------------------------------------------------------

\section{Variable Source Cluster Analysis pipeline} \label{sec:vasca}

\begin{figure}[t]
\includegraphics[width=\linewidth]{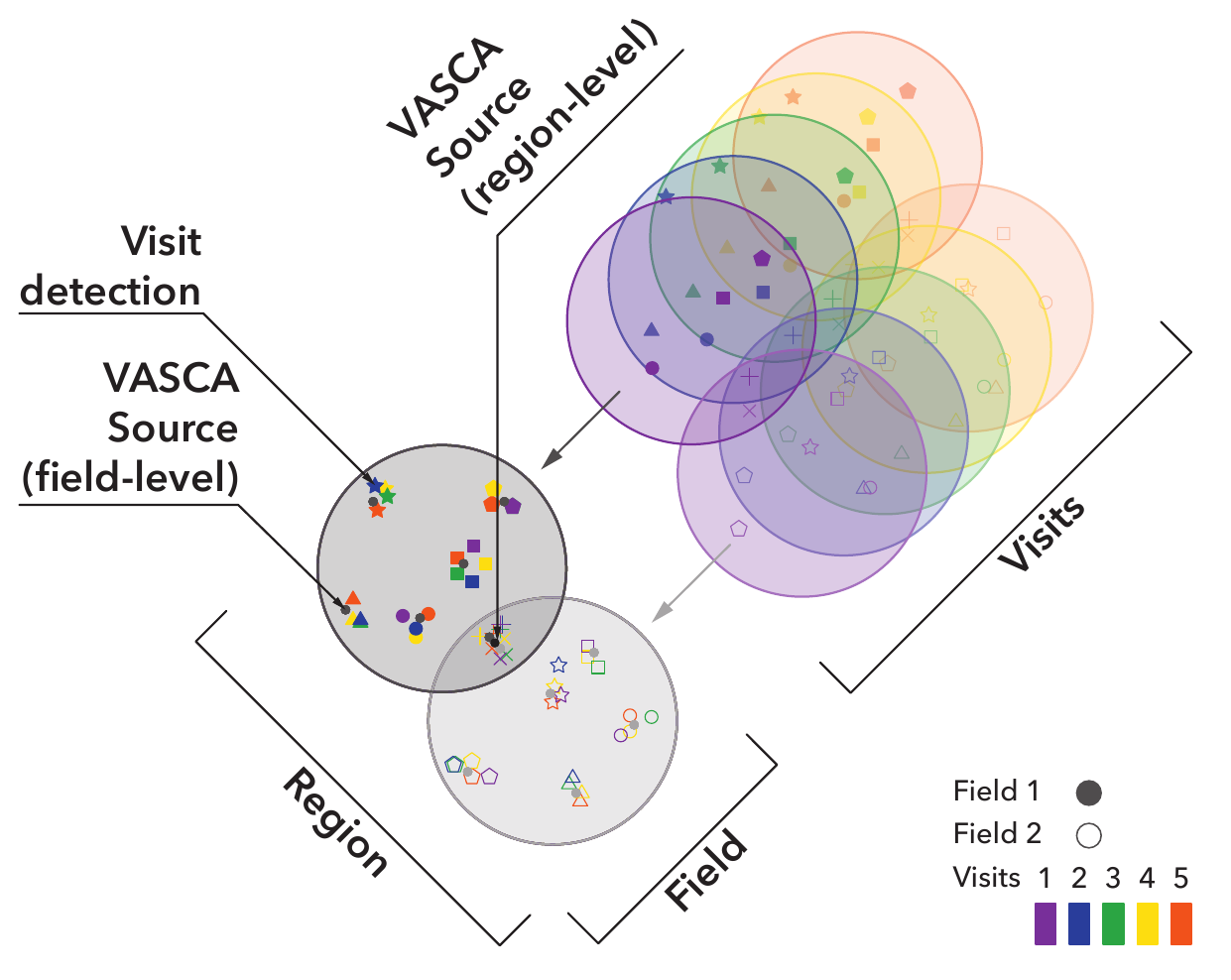}
\caption{The VASCA data model,  see main text for more details.
\label{fig:data_model}}
\end{figure}

VASCA offers a modular and scalable analysis pipeline for creating catalogs of cosmic variables from repeated photometric observations. The pipeline is implemented in an instrument independent way. We will describe it using the GALEX data set as a hands-on application in the following.

%
%-----------------------------

\subsection{Data model and GALEX data set} \label{subsec:data}

VASCA is based on a data model that describes photometric detections from repeated observations. These are referred to as ``visits'' hereafter. A set of visits observing the same patch on the sky in the same pass band defines a ``ﬁeld''.  A collection of ﬁelds deﬁnes a ``region'', as shown in figure \ref{fig:data_model}. Fields can also be overlapping, as in fact is often the case for instruments that perform surveys and follow-up observations. The data model is kept simple by only defining these three hierarchical data layers. We emphasize that observations in different pass bands and by different instruments are treated as separate fields, even if they observe the same patch of the sky. Fields are only combined on the region-level. This is fundamentally the reason for the pipeline's scaling ability and instrument independence.
%A region can be understood as the intermediary pipeline result based on which the source catalog and other science products are created.
%We emphasize, that observations of the same patch of the sky in different observing bands are considered as two separate fields in our data model.

We applied VASCA to 385 NUV-fields and 270 FUV-fields of the GALEX legacy data on the Multimission Archive at STScI (MAST). These were all fields which fullfilled the following conditions: (1) they have been visited \qty{\geq 10} times in the NUV band (2) the average NUV exposure is \qty{> 800}{\sec}. We applied our selection on the NUV band only, as FUV-only data were very rarely taken by GALEX.  The number of visits in the NUV band for each considered field in the sky is shown in the appendix \ref{sec:galexobs}.  The average number of visits for each field is 26.6 in the NUV and 18.9 in the FUV. The total exposure is \qty{12.5}{\megasec} in the NUV band and \qty{5.9}{\megasec} in the FUV band. This is approximately 7.7 times the exposure of the TDS. The applied selection ensures a rather uniform data set, with an average exposure time per visit of \qty{1220}{\sec}. The average limiting fluxes at a signal to noise of 3 for the typical visit is therefore f$_{NUV}\approx \qty{ 2.0}{\mujansky}$ and f$_{FUV}\approx \qty{ 2.2}{\mujansky}$ (m$_{AB}^{NUV} = 23.1$ and m$_{AB}^{FUV} = 23.0$, respectively).

In this work, we used the GALEX Release 6/7 data products. A description of this standard pipeline and calibration can be found online\footnote{\url{http://www.galex.caltech.edu/researcher/data.html/}} and in \citetads{2007ApJS..173..682M}. Photometry in the NUV and FUV bands are provided in these data products. The systematic flux accuracy of the pipeline photometry is expected to be 0.8\% and 2.5\% on average for the NUV and FUV bands, respectively.

%\citepads{GALEXdata}

\begin{figure}[t]
\includegraphics[width=\linewidth]{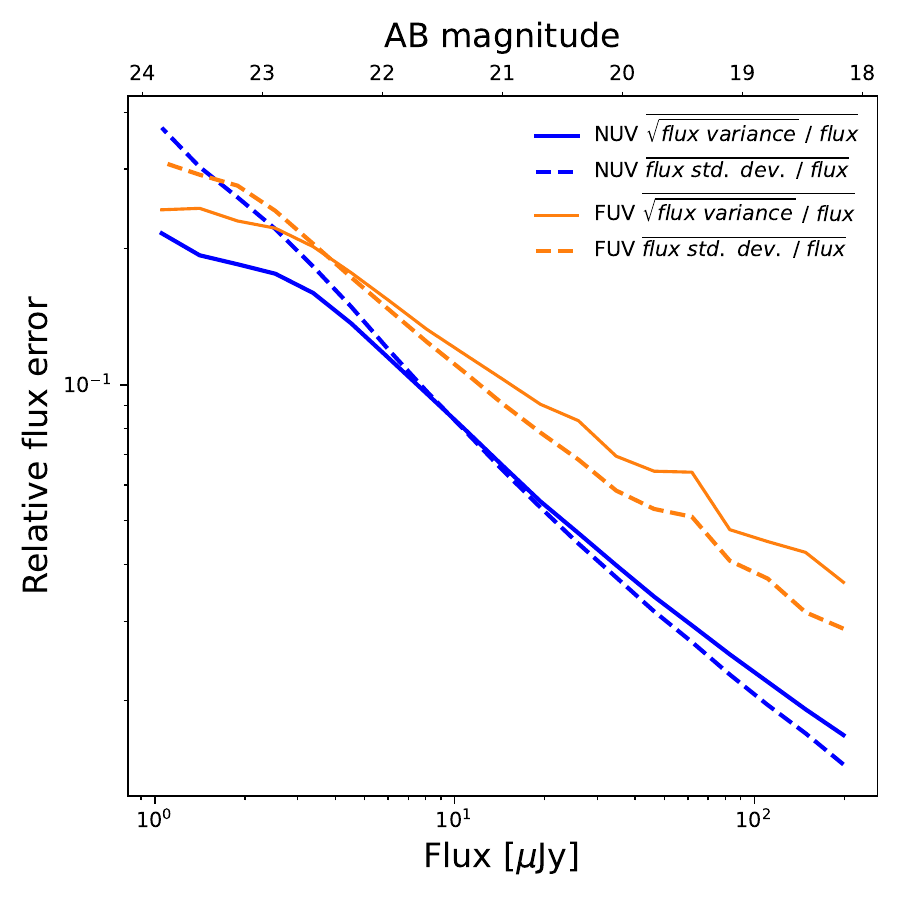}
\caption{Comparison of the observed flux variation of VASCA sources, compared to the photmetric measurement error, for the NUV (blue lines) and FUV (red lines) pass bands. See text for more details.\label{fig:fluxreco}}
%Figure from inspect_source_distributions.ipynb
\end{figure}

We cross-checked the accuracy of the photometric  measurements by comparing the observed flux variations to the ones expected for the measured flux errors. To avoid the inclusion of strongly variable sources into the data set we only considered sources with a $\chi^2$-probability to a constant flux $PVAL_{flux}>0.001$ and more than 5 light-curve points. As can be seen in figure \ref{fig:fluxreco}, the difference between the observed variations to the ones expected from the measured errors is  <10\% and <20\% in the NUV and FUV bands, respectively. Under the assumption that the sources are not time variable within the photometric sensitivity of GALEX, this comparison would yield a measure of the GALEX photometry accuracy. However, as low-level variable sources remain in the sample, this corresponds to an upper limit on the stability of the flux determination. These limits are in agreement with the systematic errors quoted previously. Note, that the discrepancy observed at low-flux levels is a selection effect: only upward-flux fluctuations are detected close to the flux sensitivity threshold.

 The GALEX images are known to contain several different artifacts, a detailed discussion can be found in \citetads{2023ApJS..268...41M}. These artifacts can potentially create artificial sources or source variability. In order to minimize their effect in our pipeline, we performed stringent quality cuts on the photometric detections. In particular, selection cuts were applied to enforce that only point-like sources are selected, as artifacts are typically extended and asymmetric. Furthermore, we restricted our analysis to the inner camera, where artifacts are more sparse. Finally, we require a minimum of three independent detections for each source, as artifacts typically do not repeat in position over multiple visits. All selection variables and values are listed in table \ref{tab:sel}.

% tab:sel
%\input{tables/vasca_sel_pars}
\begin{table*}
    \caption{
        Selection parameters used in the VASCA pipeline. \label{tab:sel}
    }
    \centering       
    \begin{tabular}{l l l }     % 3 columns 
        \hline\hline       
        \multicolumn{1}{c}{Variable} & \multicolumn{1}{c}{Description} & \multicolumn{1}{c}{Value} \\ 
        \hline
        & \multicolumn{1}{c}{\underline{Detection quality selection}}  & \\
        S2N           &	Signal to noise &	\num{>3} \\
        R$_{fov}$         &	Distance to the center of the FoV &	\qty{<0.5}{\deg} \\
        ELLIP$_{world}$   &	Ellipticity &	\num{<0.5} \\
        SIZE$_{world}$    &	Circular extension	 &\qty{<6}{\arcsec} \\
        CLASS$_{star}$	  & Extended (0) or point like (1)	 &\num{>0.15} \\
        CHKOBJ$_{type}$	  &Matched to a bright star (0=no, 1 = yes)	 &0 \\
        APPRATIO$_{flux}$ &Ratio of flux calculated with apertures of \qtylist{3.8; 6.0}{\arcsec} &	\numrange{0.3}{1.05} \\
        ARTIFACTS      & Detections on-top of variable/hot pixels and of optical reflections are ignored & \numlist{2;4;8;128;256} \\
        \hline
         & \multicolumn{1}{c}{\underline{Variable source selection}}  & \\
        FLUX	  & Spectral flux density	 & \qty{0.145}{\mujansky} to \qty{575.4}{\mujansky}\\
        N$_{det}$           &	Number of detections &	\num{>3} \\
        PVAL$_{flux}$    &	Probability of constant flux	 &\num{<5.73e-7} \\
        NXV$_{flux}$	  & Flux normalized excess variance	 &\num{>0.001} (\num{>0.01}) \\
        FRATIO$_{co}$ & Ratio of the mean flux to the co-add flux& \num{>2}\\
        S2N$^{fdiff}_{co}$ & Signal to noise of the flux to co-add flux difference& \num{>7}\\
        QVAL$_{pos}$         &	Cluster position quality parameter & \num{<5.73e-7}\\
        XV$_{pos}$  &	Positional excess variance & \num{<2} arcsec$^{2}$ \\
        \hline                  
    \end{tabular}
    \tablefoot{More details on the detection variables can be found in the GALEX documentation\url{http://www.galex.caltech.edu/wiki/}. Selection parameters are typically equal for the NUV and FUV bands. If the FUV parameter value differs, it is listed in brackets.}
\end{table*}

%
%-----------------------------

\subsection{Pipeline and variability selection} \label{subsec:vasca_pipeline}

A schematic representation of the analysis flow of VASCA is shown in the appendix \ref{sec:processing}. Photometric detections of each visit are inputs to the pipeline. After quality selection, detections are clustered for all visits of each field using the MeanShift algorithm (\citeads{2011JMLR...12.2825P}; \cite{MeanShift1000236}). The clustering bandwidth is always fixed to \qty{4}{\arcsec}, significantly larger than the typical absolute astrometric performance of \qty{\lessapprox 1.5}{\arcsec} of GALEX \citepads{2007ApJS..173..682M}.

To handle different observation filters and overlapping fields, a second clustering is performed for all clusters obtained in the field analysis. The clusters obtained in this second step define the position of a 1UVA source. Its position and mean flux are calculated as the error-weighted mean from all detections associated to the individual cluster. For GALEX observations, additional source photometry is typically provided for the sum of all visit images in a field, the so called co-add image. This information is also fed into the pipeline. Co-add image detections are clustered and associated to the sources previously derived from the visit detections. 

Statistical measures are calculated to diagnose the source time variability. The primary statistic is the $\chi^2$-probability for a constant flux. We select sources which are incompatible  with a constant flux at a 5-$\sigma$ level. We also select sources which have a significant difference between the mean flux and the co-add flux. The former is defined as the error-weighted mean of the flux in all visit-level detections, while the latter is the one obtained from the photomety of the co-add images. This cut is applied to select sources that are only detected during bright flaring periods in a few visits. All selection values are listed in table \ref{tab:sel}. To take into account possible systematic flux variations we also select on the normalized excess variance of the flux $NXV_{flux} = (Var_{flux} - \overline{err_{flux}^2}) /  \overline{flux}^2$, where $Var_{flux}$ and $err_{flux}$ is the variance and error of the flux measurements \citepads{2003MNRAS.345.1271V}. 
 The selection on the minimum excess variance corresponds to a flux variability of $\approx$ 3\% and $\approx$ 10\% for the NUV and FUV band, respectively. This is \num{\approx 4} times the expected photometric stability of GALEX discussed in the previous section, making sure that no instrumental variations lead to artificial variability.

%
%-----------------------------

\subsection{Source association} \label{subsec:vasca_src}

In order to find multi-frequency counterparts we check for positional coincidences within \qty{1.5}{\arcsec} for all 1UVA sources. This match is done with all sources listed in the SIMBAD database \citepads{2000A&AS..143....9W}, the \textit{Gaia}-DR3 and WD catalogs (\citeads{2023A&A...674A...1G}; \citeads{2021MNRAS.508.3877G}) and the recent GALEX Flare Catalog \citepads{2023ApJS..268...41M}. The latter lists sources which are variable within one visit in GALEX data.  If multiple counterparts are found in one catalog, the closest one is preferred. 

In order to obtain a Spectral Energy Distribution (SED), we use the VizieR photometry tool\footnote{\url{http://vizier.cds.unistra.fr/vizier/sed/doc/}} . It provides all SED points from all entries in the VizieR database \citepads{2000A&AS..143...23O}. We caution that the latter is done without specific checks on the quality of these catalogs.  Finally, we also query if SDSS spectra are available for each source \citepads{2022ApJS..259...35A}.

\begin{figure}[]
\includegraphics[width=\linewidth]{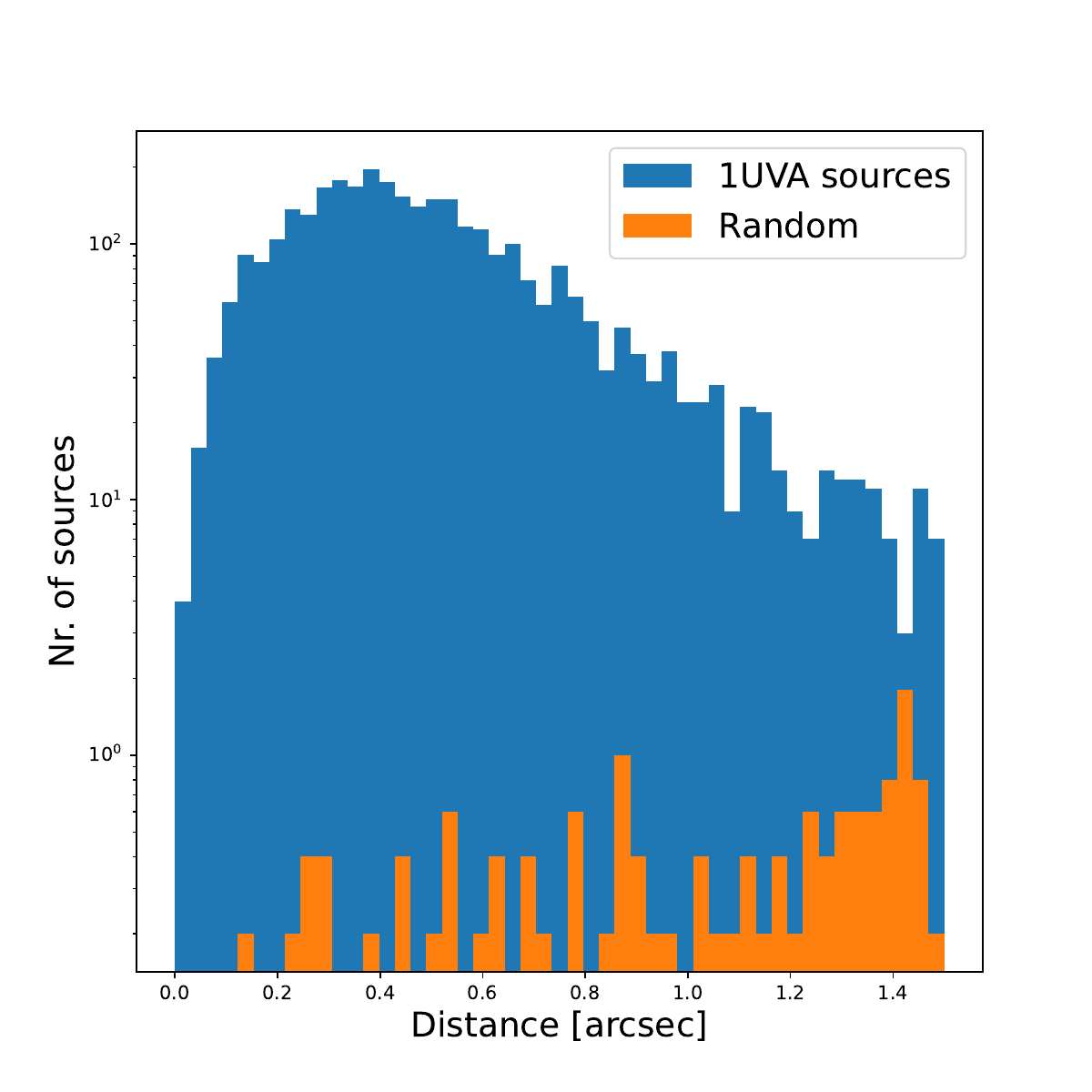}
\caption{Angular distance between the 1UVA sources and the associated \textit{Gaia}-DR3 sources, for the measured source positions (blue bars) and randomly scattered positions (orange bars).
\label{fig:ang_dist}}
\end{figure}

%
%-----------------------------

\subsection{Periodicity search} \label{subsec:vasca_period}

We performed a periodicity search using a Lomb-Scargle periodogram (\citeads{1976Ap&SS..39..447L}; \citeads{1982ApJ...263..835S}) in the frequency range of \qty{0.03}{\per\day} to \qty{2}{\per\day}. For all 1UVA sources we estimate significance of the main peak in the periodograms with the method of \citetads{2008MNRAS.385.1279B}. We include periodicity frequency peaks with a significance greater than 4-$\sigma$ in the catalog, if the light curve contained more than 20 points. We are aware that the usage of these periodicity detections methods has significant caveats in the case of sparse binning,  as is typically the case for the light curves in the catalog (for a discussion see e.g. \citetads{2018ApJS..236...16V}). We therefore consider the reported periodicities only as tentative. They can be useful to search for periodicity for these sources in the future, with more evenly sampled light curves. 

%
%-------------------------------------------------------------------

\section{Catalog of ultraviolet variable sources} \label{sec:catalog}

In total, our pipeline found 1991105 UV-sources. Out of these, 4202 sources pass the flux variability selection. The latter compose the 1UVA catalog. A list of the information available for each source in the catalog is provided in appendix \ref{sec:example_src}. On average, the light curves of 1UVA sources contain 6.1 photometric measurements in the NUV and 3.4 in the FUV. A wide range of time scales is probed, from \qty{\approx 90}{\minutes} to \qty{\approx 8}{\years}. The distribution of time differences between light-curve points is shown in the appendix \ref{sec:galexobs}.

We find multi-frequency counterparts for 3656 sources: 3301 sources have a counterpart in the a \textit{Gaia}-DR3 catalog and 2686 sources a counterpart in the SIMBAD database. The distance distribution between the 1UVA positions to the associated \textit{Gaia}-DR3 sources is shown in figure \ref{fig:ang_dist}. The average angular distance is \qty{0.40}{\arcsec}. As the positional uncertainties of \textit{Gaia}-DR3 sources are negligible compared to the ones of GALEX, this also corresponds to the mean positional accuracy of the 1UVA sources.

In order to check the chance probability of false associations, we shifted the 1UVA source positions randomly between \qtylist{2;60}{\arcsec} five times and performed the source association again: on average, only 14.8 \textit{Gaia}-DR3 associations are found, their distance distribution is also shown in figure \ref{fig:ang_dist}. The low number of random matches confirms the expectation that only few sources are wrongly associated in the catalog. This also confirms that, if at all, only a few spurious sources are part of the 1UVA catalog.

%
%-----------------------------

\subsection{Source classes} \label{subsec:classes}

\begin{figure}[t]
\includegraphics[width=\linewidth]{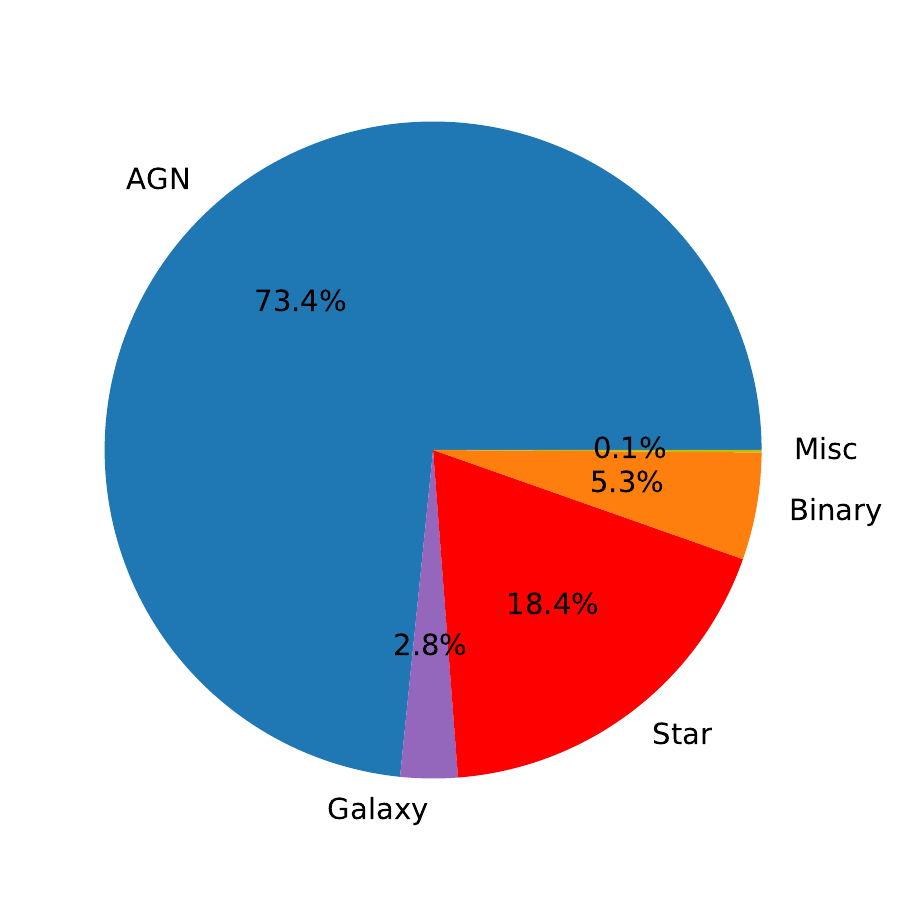}
\caption{Object groups for 1UVA counterpart sources with a secured source type in the SIMBAD database. A more detailed subdivision is given in table \ref{tab:ogroups}.
\label{fig:ogroups}}
\end{figure}

\begin{figure*}[ht]
\centering
\includegraphics[width=0.48\linewidth]{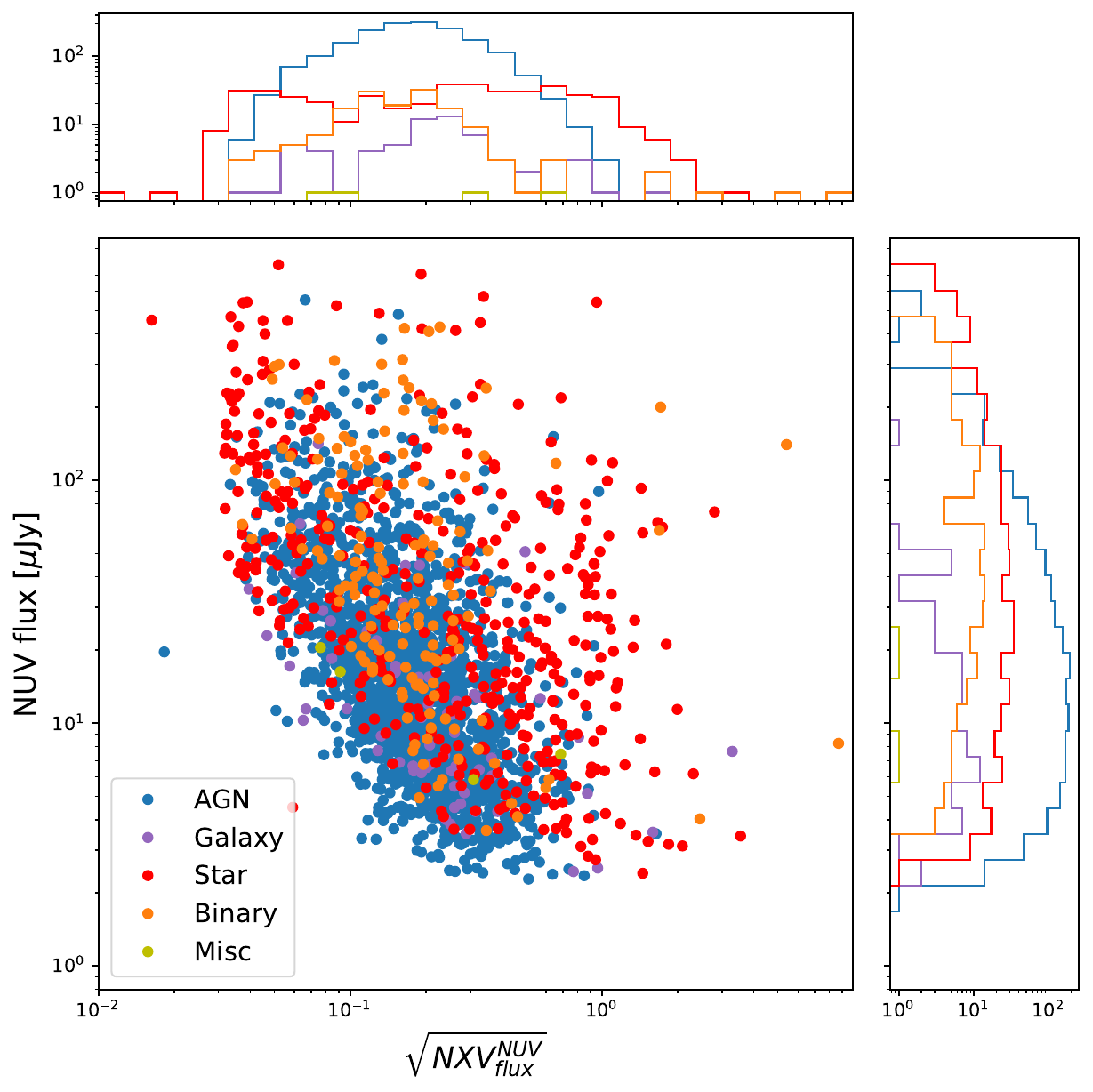}
\hspace{.02\textwidth}
\includegraphics[width=0.48\linewidth]{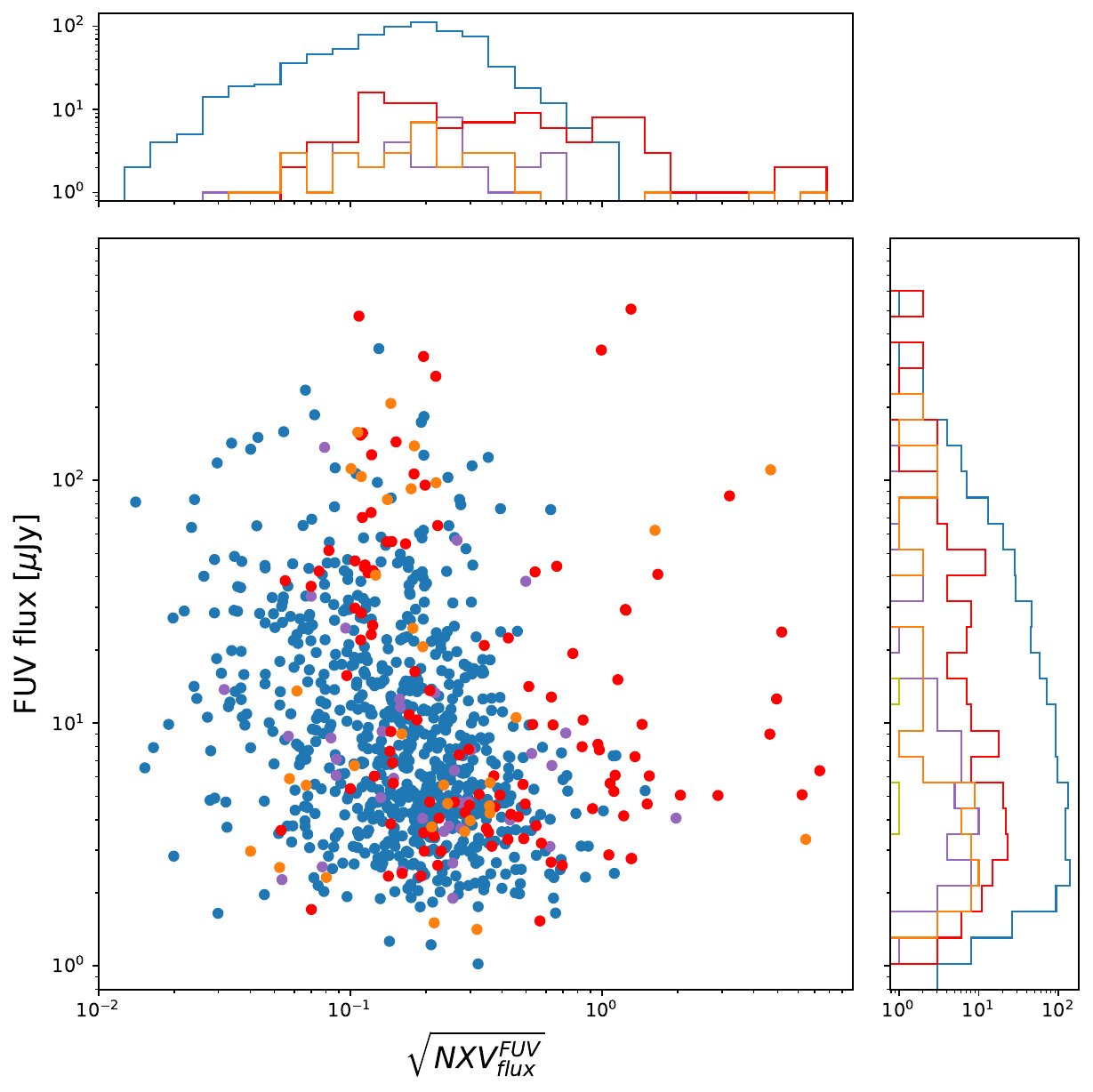}
\caption{Flux versus the relative amplitude of flux variations, given by the square root of the normalized excess variance, as a function of the mean flux, for the NUV (left) and FUV (right) pass bands. The axes range is the same in both bands, for an easier comparison. The sub-panels at the top and right of each image show the total number of sources projected onto the respective axis. \label{fig:nxv}}
%Figure from inspect_source_distributions.ipynb
\end{figure*}

To classify the types of the 1UVA sources, we rely on the classification used in the SIMBAD data base\footnote{\url{http://vizier.u-strasbg.fr/cgi-bin/OType?$1}}. The distribution of source types is shown in figure \ref{fig:ogroups} and listed in more detail in table \ref{tab:ogroups}. Both of them only show SIMBAD counterpart sources with a secured source type. 

In general, a large diversity of sources is found.  As expected, the vast majority of sources are Active Galactic Nuclei (AGN, \qty{\approx 73}{\percent}). Among these, the subclass of quasars dominated the sample. The second largest class of sources are stars, either single (\qty{\approx 18}{\percent}) or in binary systems (\qty{\approx 5}{\percent}). Non-active Galaxies (\qty{\approx 2.8}{\percent}) and one HII region are also found; these large diffuse objects must have a variable source within them, which is of unknown type at this point.

In two cases, 1UVA sources are associated to a Supernova (SN) explosion. For the first case, 1UVA~J141829.9+534331.0, the time profile corresponds to the SN~PS1-11pf \citepads{2015ApJ...799..208S}. In the second case, 1UVA~J33308.1-271452.5, the observed UV variability precedes the associated ``SN~cdfs1~r~20121007~43A'' by several years. In addition, the latter is only classified as "Probably Supernova" in the original catalog \citepads{2015A&A...584A..62C}. This source is therefore very likely not a SN.

The square root of the flux-excess variance of the different source classes is shown in figure \ref{fig:nxv}, for the NUV and FUV bands. In both bands, the observed variability amplitude varies between a few percent to a factor $\approx 10$. The variability amplitude is generally larger and extends to higher values for stellar objects than for AGN. A similar trend had already been found in the TDS survey \citepads{2013ApJ...766...60G}.

Due to the richness of the data set, a deeper study of all sources types found in the 1UVA catalog is beyond the scope of this article. Here, we will focus on two findings: first, the large variety of different stars that are found in the 1UVA catalog. Second, two WDs are found to be variable, even though their SED show no sign of any companion star. We will go into more details on both of these findings in the following sections.

% tab:ogroups
%\input{tables/simbad_types}
\begin{table*}
    \caption{
        Types of 1UVA counterparts in the SIMBAD database.
        \label{tab:ogroups}
    }
    \centering       
    \begin{tabular}{l l l l}
        \hline\hline       
        \multicolumn{1}{c}{Object group} & \multicolumn{1}{c}{Type} & \multicolumn{1}{c}{Description}& \multicolumn{1}{c}{Nr. sources} \\
        \hline
        AGN & AGN & Active Galaxy Nucleus & 87 \\
        AGN & BLL & BL Lac & 11 \\
        AGN & Bla & Blazar & 2 \\
        AGN & QSO & Quasar & 1620 \\
        AGN & Sy1 & Seyfert 1 Galaxy & 127 \\
        AGN & Sy2 & Seyfert 2 Galaxy & 2 \\
        AGN & SyG & Seyfert Galaxy & 1 \\
        AGN & rG & Radio Galaxy & 1 \\
        Binary & ** & Double or Multiple Star & 2 \\
        Binary & CV* & Cataclysmic Binary & 11 \\
        Binary & EB* & Eclipsing Binary & 109 \\
        Binary & No* & Classical Nova & 2 \\
        Binary & SB* & Spectroscopic Binary & 9 \\
        Galaxy & BiC & Brightest Galaxy in a Cluster (BCG) & 2 \\
        Galaxy & EmG & Emission-line galaxy & 2 \\
        Galaxy & G & Galaxy & 62 \\
        Galaxy & GiC & Galaxy towards a Cluster of Galaxies & 3 \\
        Galaxy & GrG & Group of Galaxies & 2 \\
        Misc & HII & HII Region & 1 \\
        Misc & SN* & Supernova & 2 \\
        Star & * & Star & 157 \\
        Star & AB* & Asymptotic Giant Branch Star & 1 \\
        Star & BS* & Blue Straggler & 1 \\
        Star & BY* & BY Dra Variable & 1 \\
        Star & Em* & Emission-line Star & 1 \\
        Star & Er* & Eruptive Variable & 8 \\
        Star & GlC & Globular Cluster & 1 \\
        Star & HB* & Horizontal Branch Star & 5 \\
        Star & Ir* & Irregular Variable & 2 \\
        Star & LM* & Low-mass Star & 4 \\
        Star & LP* & Long-Period Variable & 3 \\
        Star & PM* & High Proper Motion Star & 20 \\
        Star & Pe* & Chemically Peculiar Star & 1 \\
        Star & Pu* & Pulsating Variable & 12 \\
        Star & RG* & Red Giant Branch star & 2 \\
        Star & RR* & RR Lyrae Variable & 203 \\
        Star & Ro* & Rotating Variable & 5 \\
        Star & TT* & T Tauri Star & 1 \\
        Star & V* & Variable Star & 22 \\
        Star & WD* & White Dwarf & 5 \\
        Star & WV* & Type II Cepheid Variable & 1 \\
        Star & Y*O & Young Stellar Object & 1 \\
        Star & cC* & Classical Cepheid Variable & 2 \\
        Star & dS* & delta Sct Variable & 4 \\
        Star & s*b & Blue Supergiant & 1 \\
        \hline                  
    \end{tabular}
\end{table*}
%
%-----------------------------

\subsection{UV-variable stars}

\begin{figure}[t]
\includegraphics[width=\linewidth]{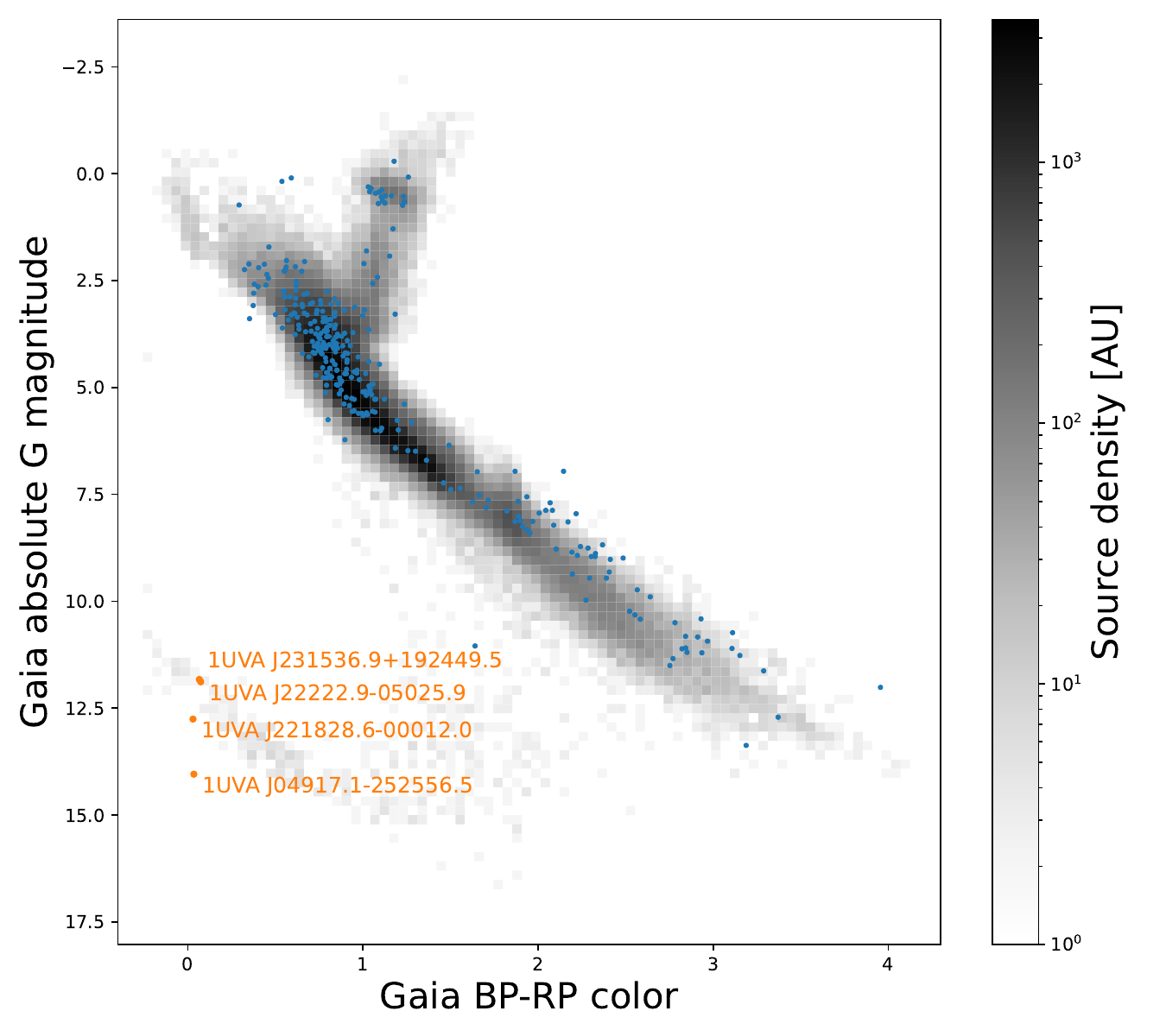}
\caption{Hertzsprung-Russel Diagram for VASCA sources with \textit{Gaia}-DR3 counterparts (blue points, see text). The gray background map shows the source density of randomly selected \textit{Gaia}-DR3 sources for comparison. The \textit{Gaia} source selection cuts are the same for both cases. The orange marker shows the UV-variable WDs discussed in section \ref{subsec:catalog_wd}.
\label{fig:HR}}
\end{figure}

As can be seen in table \ref{tab:ogroups}, many different stellar classes are found to be UV-variable. The dominant class is RR Lyrae stars, as expected from previous studies \citepads{2013ApJ...766...60G}.  Several other pulsating stars are also found, as Pulsating Variables, Cepheids and delta Sct Variables. Perhaps more surprisingly, 20 High Proper Motion stars are found in the sample. 

Using the \textit{Gaia}-DR3 associations, we constructed the Hertzsprung-Russel (HR) diagram of \textit{Gaia} counterparts of 1UVA sources, shown in figure \ref{fig:HR}. The figure also shows the star density for a random sample of \textit{Gaia} sources. For both, we applied the same quality cuts on the \textit{Gaia} measurements: the signal to noise of the blue and red filter photometry and the parallax measurements all have a signal to noise greater than 10. The fluxes of sources beyond \qty{150}{\parsec} are corrected for dust extinction using the estimates from the low-resolution \textit{Gaia} spectra \citepads{2023A&A...674A..27A}. Only sources for which the latter extinction estimate was available were included in the sample. In addition, sources were only included if they have an extinction in the G~band $A_G < 1$. Sources closer than \qty{150}{\parsec} are all included and not corrected for extinction, as the effect is expected to be small in this case $A_G \lesssim 0.02$.

It is interesting to note that 1UVA sources are found throughout the HR diagram, even though there is strong selection bias towards bluer stars due to the UV selection of the sample. UV variability seems to be ubiquitous for stars, independent whether they are on the main sequence or in the horizontal branch. In the former, a shift towards redder and brighter sources is seen for stars with absolute magnitudes larger than \num{\approx7}. This is likely due to the fact that UV variable sources are preferably binary systems \citepads{2018A&A...616A..10G}. Finally, WDs stars are also found in the sample. 

%
%-----------------------------

\subsection{UV-variable White Dwarfs} \label{subsec:catalog_wd}

% tab:sigleWDs
%\input{tables/white_dwarfs}
\begin{table*}
    \centering 
    \caption{
        Properties of the WDs associated with the sources shown in figure \ref{fig:wd_sed_lc}.
        \label{tab:sigleWDs}
    }
    \begin{tabular}{l | l l l l l}
        \hline \hline
        1UVA ID & J04917.1-252556.5 & J221828.6-00012.0 & J22222.9-05025.9 & J231536.9+192449.5 \\
        WD ID & J004917.14-252556.81 & J221828.58-000012.17 & 	J022222.85-005026.59 & J231536.88+192449.14\\
        Distance & 99.6 pc & 121.7 pc& 371.6 pc& 168.9 pc\\
        %Parallax& 39.4 & 48.4 & 7.8 & 39.2 \\
        Spectrum& DA & DAH & DA & Unknown\\
        Temperature& 14145 K & 11514 K & 11831 K & 12483 K\\
        \hline
    \end{tabular}
    \tablefoot{The WD ID refers to the \textit{Gaia}-EDR3 WD catalog and the temperature to the Black Body fit shown in figure \ref{fig:wd_sed_lc}.}
\end{table*}

WDs are known to be strongly variable in the ultra-violet when they have a stellar companion. The accretion of matter of the companion star can lead to strong novae explosions in Cataclysmic Variables (CVs) \citepads{2023MNRAS.524.4867I}. Indeed, 11 CVs counterparts are found for 1UVA sources in the SIMBAD database, see table \ref{tab:ogroups}. Time variability has also been found for WDs with sub-stellar companions: the obscuration during eclipses and the heating of the companion can cause flux periodicity on time scales between a few hours and several days (\citeads{2016Natur.533..366H}; \citeads{2021ApJ...919L..26V}). On short time scales of \qty{\approx 10}{\min}, periodic UV variability has been found for isolated WDs from their rotating photosphere, in so called Pulsators \citepads{2019MNRAS.486.4574R}.  

% \begin{figure*}[t]
% %\plotone{LC_357455.pdf}
% \includegraphics[width=\linewidth]{figures/LC_357455.pdf}
% \caption{Light curve for the sources 1UVA J04917.1-252556.5. The dashed line shows the average NUV flux for comparison.  Properties of the multi-wavelength counterpart WD~J004917.14-252556.81 are listed in table \ref{tab:sigleWDs}. \label{fig:wd_lc}}
% \end{figure*}

\begin{figure*}[h]
\centering
\includegraphics[width=0.9\linewidth]{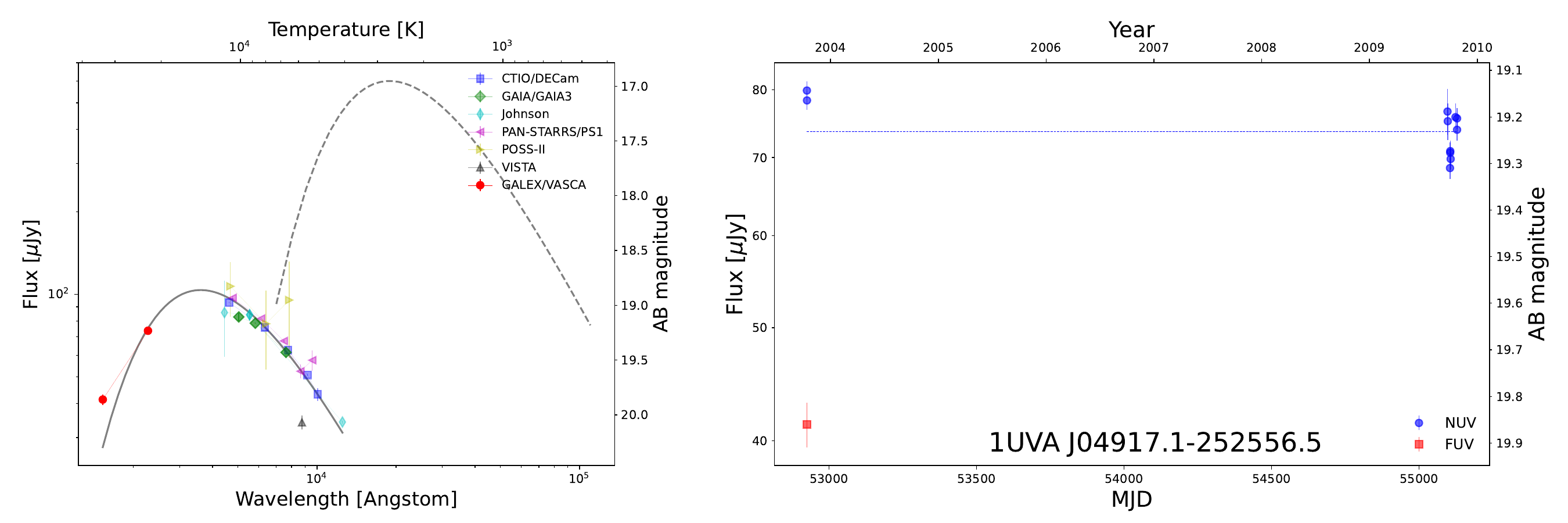}
\includegraphics[width=0.9\linewidth]{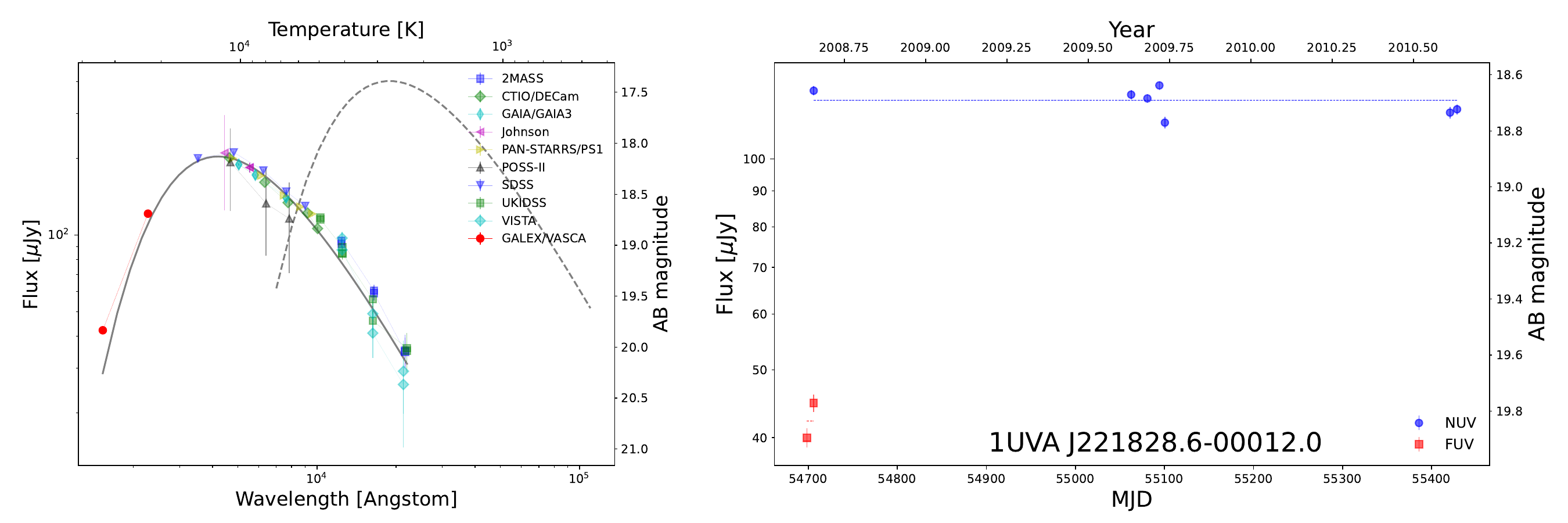}
\includegraphics[width=0.9\linewidth]{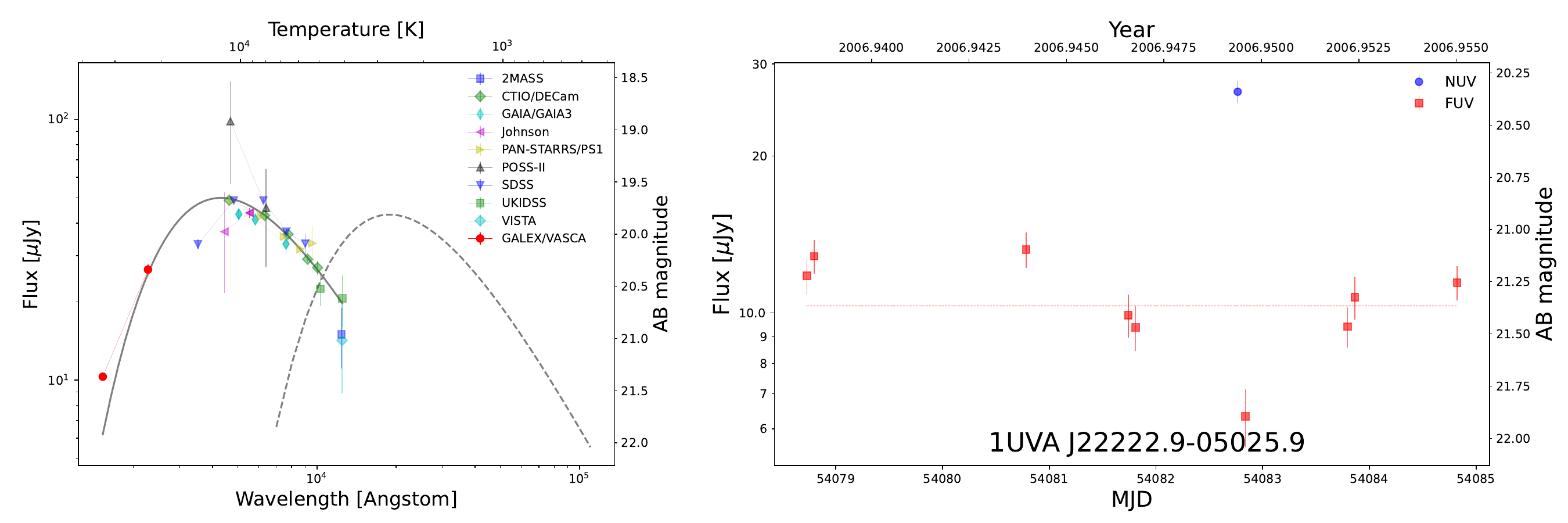}
\includegraphics[width=0.9\linewidth]{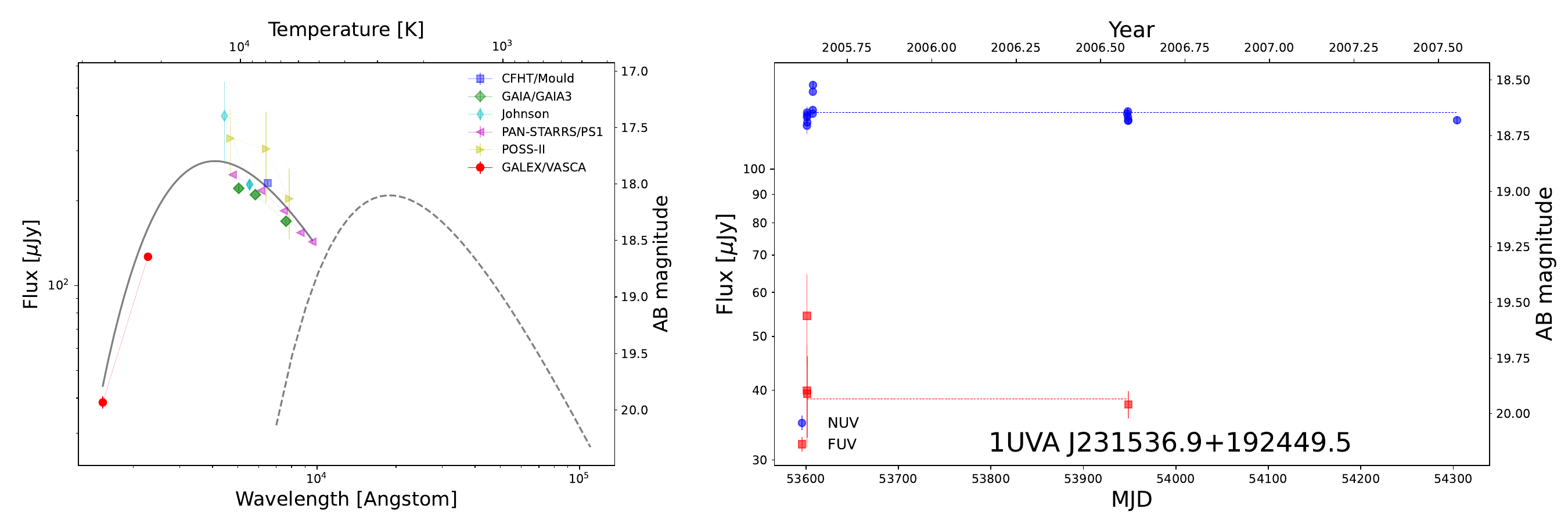}
\caption{SED (left) and light curve (right)  for the sources listed in table \ref{tab:sigleWDs}. \textbf{Left:} The straight line shows the best fit to a black body emission spectrum. The black body temperature is listed in table \ref{tab:sigleWDs}. The dashed line shows the SED of a brown dwarf star in a black body approximation (see main text). \textbf{Right:} The dashed line shows the mean flux value.\label{fig:wd_sed_lc}}
\end{figure*}

We searched for isolated variable WDs among the 1UVA sources. For this, we selected sources which have a counterpart that is classified as a WD at \qty{>90}{\percent} confidence in the \textit{Gaia}-EDR3 WD catalog. Four 1UVA counterpart sources passed this selection, their properties are listed in table \ref{tab:sigleWDs}.  Note, that this classification is more reliable than the one of the SIMBAD database used previously. Indeed, two sources which had been classified as WDs in the SIMBAD database showed a probability $< 35\% $ of being a WD in the \textit{Gaia}-EDR3 WD catalog (1UVA~J234829.1-92500.3 and 1UVA~J221409.9+05246.0). We therefore did not include them in this discussion.

The SED of these sources is shown in the left panels of figure \ref{fig:wd_sed_lc}. No companion star is visible in all cases. To emphasize this point, we show the SED of a putative dim brown dwarf star companion in the SED in this figure. The star was assumed to have a radius of 0.1\(R_\odot\) and to have a temperature of 2700K. As one can see, the measured fluxes are about an order of magnitude below the expected stellar emission, particularly in the infrared.

The UV light curve of the four WDs is shown in the right panels of figure \ref{fig:wd_sed_lc}. All of them are clearly variable. The variability time scale has to be larger than the typical observation time of $\approx \qty{24}{\min} $. In several cases, indications of long-term trends of time scales of days to years are apparent: e.g. for WD~J004917.14-252556.81 all fluxes measured in 2009 are below the ones measured in 2004. However, due to the scarcity of the data, it is possible that these trends are in fact due to the random sampling of shorter term variability.

One of the sources, WD J221828.58-000012.17, is classified as a magnetic WD.  Flux variations from magnetic WDs are well known \citepads{2013ASPC..469..429L}: they are related to the rotation of the WD and have a period between several minutes to several days. The typical peak-to-peak amplitude of the flux variations are a few percent. Even so no periodicity can be measured in the UV data due to its scarcity, this might explain the observed flux variations for this source.

Two sources, WD~J004917.14-252556.81 and WD~J022222.85-005026.59, are classified as normal DA WDs. Variability time scales $\gtrsim \qty{24}{\min} $ is not usual for these sources. It might hint at an ongoing accretion from an undetected substellar companion. However, no companion was seen in the SED and no spectral lines from an accretion disk can be seen in the optical to infrared spectrum \citepads{2023MNRAS.518.2341K}. Another possibility is that planetary debris is absorbing the light from the WD periodically \citepads{2020ApJ...897..171V}. A third possibility might be that the temperature of the photosphere is changing due to a yet unknown reason.

Particularly interesting is the WD~J004917.14-252556.81, as pulsations with a period of $T_{P1} = \qty{221.36}{\sec}$ and $T_{P2} = \qty{209.3}{\sec}$ have been found from this source \citepads{2023MNRAS.522.2181K}. The peak-to-peak amplitude of the pulsations is \qty{\approx30}{\percent}. This source is the most massive WD from which pulsations have been found to date. We looked for these periodic signals in GALEX data: we derived a light curve sampled in \qty{40}{\sec} bins using the gPhoton tool \citepads{2016ApJ...833..292M}. We found this to be the smallest time binning at which the photon count rate per bin is still acceptable. Note, that this time binning does unfortunately not allow to resolve the periods $T_{P1}$ and $T_{P2}$ separately. 

As the WD oscillations might drift with time \citepads{2023MNRAS.522.2181K}, we searched for periodicity's running a Lomb Scargle test for the four observing blocks in the GALEX data: MJD 52925.31975--52925.61277, 55096.10218--55097.06439, 55104.57558--55106.78567 and 55123.12991--55128.62543. The periodogram is shown in the appendix \ref{sec:puls}. No significant pulsations were found in all except the third observing block. During this observing block a peak is found in the periodogram at $T_{P} = \qty{\approx 218}{\sec}$ with a false probability of \qty{0.10}{\percent}. This period is in agreement with $T_{P1}$ and $T_{P2}$ within the accuracy of the time binning of our data. We therefore consider this an indication that the found oscillations are also present in the UV, at least during some time periods. More sensitive UV observations with a finer time resolution will be needed to settle this question.
%
%-------------------------------------------------------------------

\section{Summary and Outlook} \label{sec:summary}

We have presented the 1UVA source catalog of variable UV sources. We described a novel analysis pipeline, called VASCA, to cluster and diagnose sources found in photometric data. We applied the VASCA pipeline to GALEX data, finding 4202 variable UV-sources that vary in time scales between \qty{\approx 30}{\min} to several years. We found a multi-frequency counterpart for 3655 of these sources.  As expected, AGN dominate the source sample by numbers.  The second largest group are variable stars.

We found, that UV-variability is ubiquitous for stellar objects, even for those which are not in binary systems; UV-variable stars are found in all regions of the HR diagram.  We then focused our attention on WDs, which are not in CV systems, finding four variable WDs. One of them, WD~J004917.14-252556.81, is particularly interesting. This source has recently been found to be the most massive WD with seismic periodic oscillations \citepads{2023MNRAS.522.2181K}. We found indications for these pulsations also in the GALEX UV data. The observed UV variability from this source is puzzling and we speculated on several possible scenarios.

Due to its modularity and instrument independence, VASCA can be applied to different surveys in the future. The only requirement is that photometric measurements are available on a visit level, before co-adding the data. Such data is for instance expected to be available for the upcoming ULTRASAT mission \citepads{2023arXiv230414482S}, \textit{Vera Rubin} telescope \citepads{2019ApJ...873..111I} and Cherenkov Telescope Array \citepads{2011ExA....32..193A} data. For future UV data, the 1UVA catalog provides a long time baseline, that can be used to study UV flux variability over several decades.

Finally, we want to mention that due to richness of the 1UVA dataset, we could only focus our attention on selected sources in this work. We encourage the usage of the VASCA code and the catalog data products for further studies.

%
%-------------------------------------------------------------------

\begin{acknowledgements}
We thank Thomas Kupfer for insightful discussion on the presented work. This work made use of Astropy \citep{2022ApJ...935..167A}, scikit-learn \citep{2011JMLR...12.2825P}, the SIMBAD database \citep{2000A&AS..143....9W} and the VizieR catalogue access tool (DOI : 10.26093/cds/vizier). 
\end{acknowledgements}

\bibliographystyle{aa}
\bibliography{vasca}

%
%-------------------------------------------------------------------
\onecolumn
\begin{appendix}
\FloatBarrier
\section{GALEX observations} \label{sec:galexobs}

\begin{figure*}[h]
\centering
\includegraphics[width=0.9\linewidth]{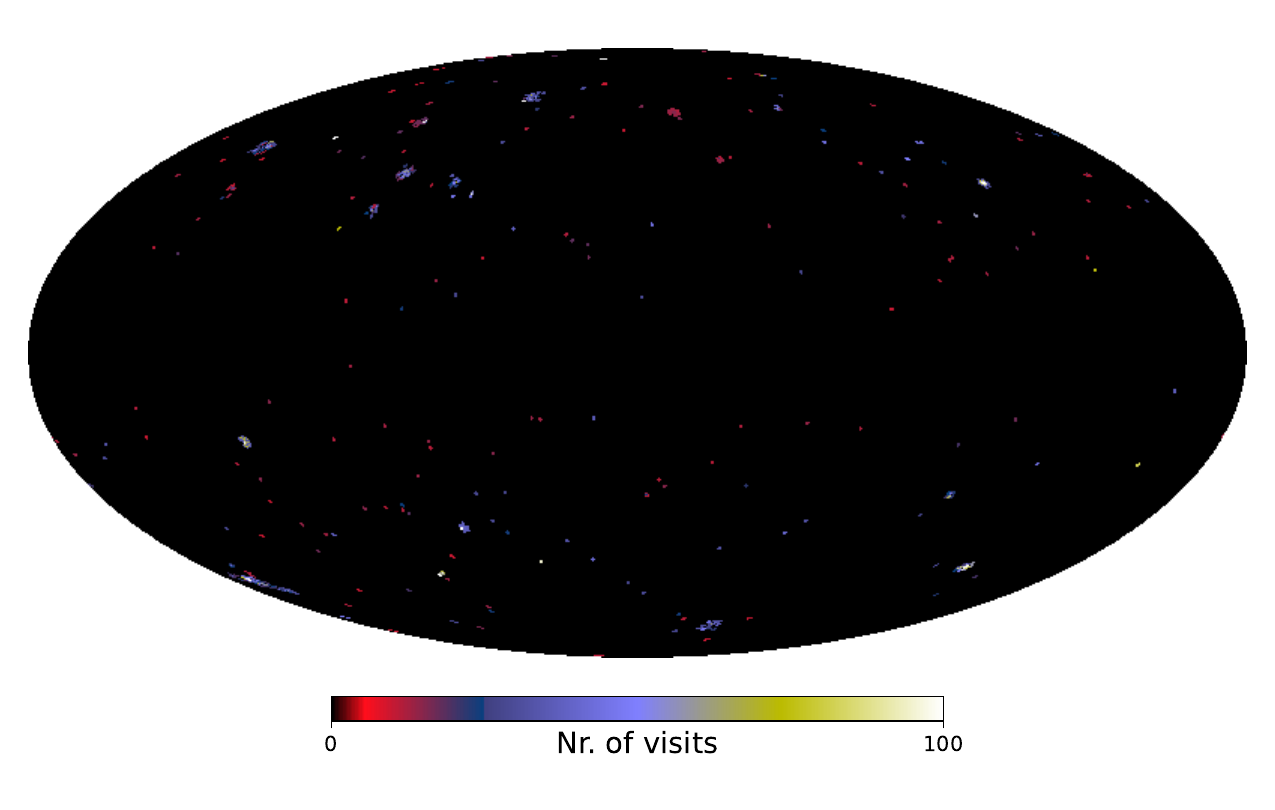}
\caption{Number of visits with NUV exposure for each field considered in the 1UVA catalog. The sky map is shown in galactic coordinates in a Mollweide projection.
\label{fig:sky_nr_vis}}
%Figure from inspect_GALEX.ipynb
\end{figure*}

\begin{figure}[h]
\centering
\includegraphics[width=0.6\linewidth]{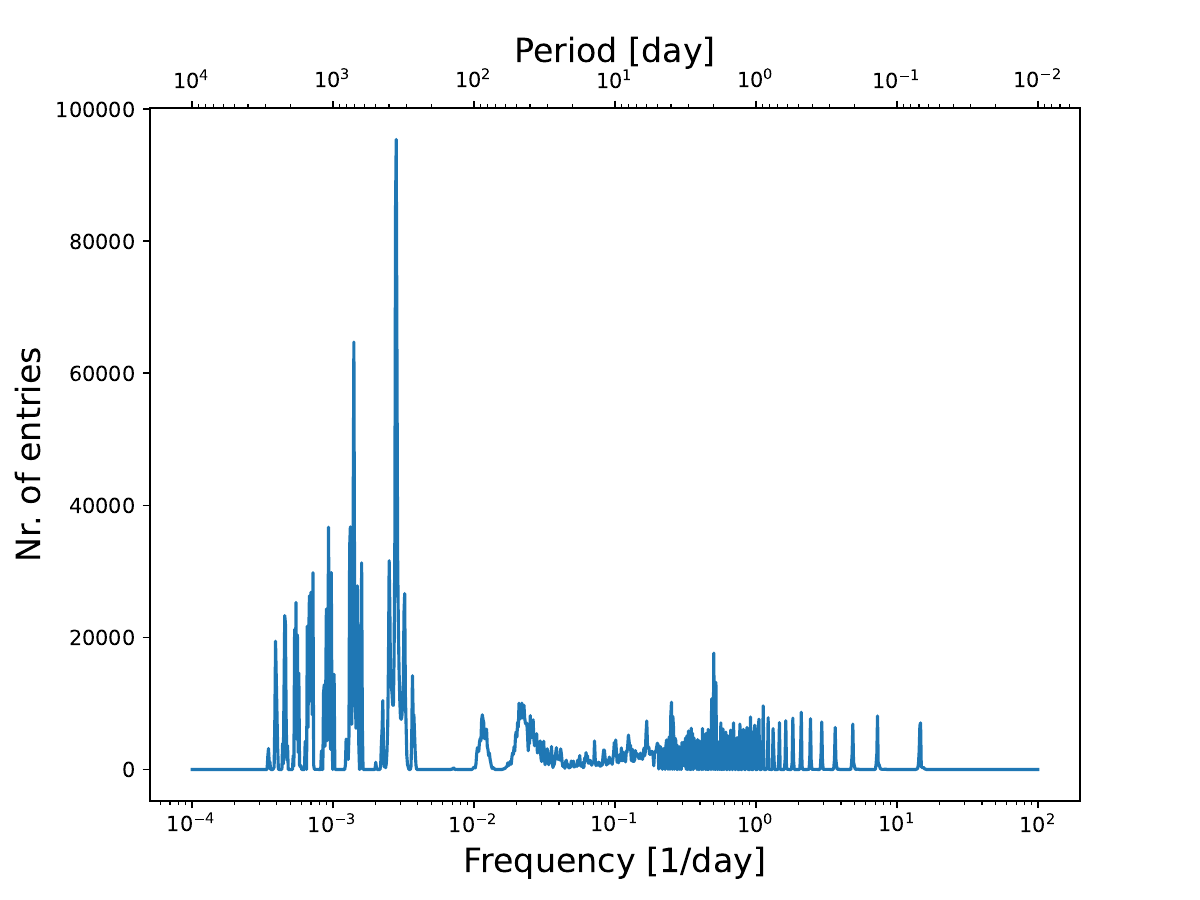}
\caption{Time difference distribution between all combinations of light-curve points for the 1UVA sources.
\label{fig:dtime}}
\end{figure}

\FloatBarrier
\clearpage
\section{VASCA processing flow} \label{sec:processing}

\begin{figure}[h]
\centering
\includegraphics[width=0.45\linewidth]{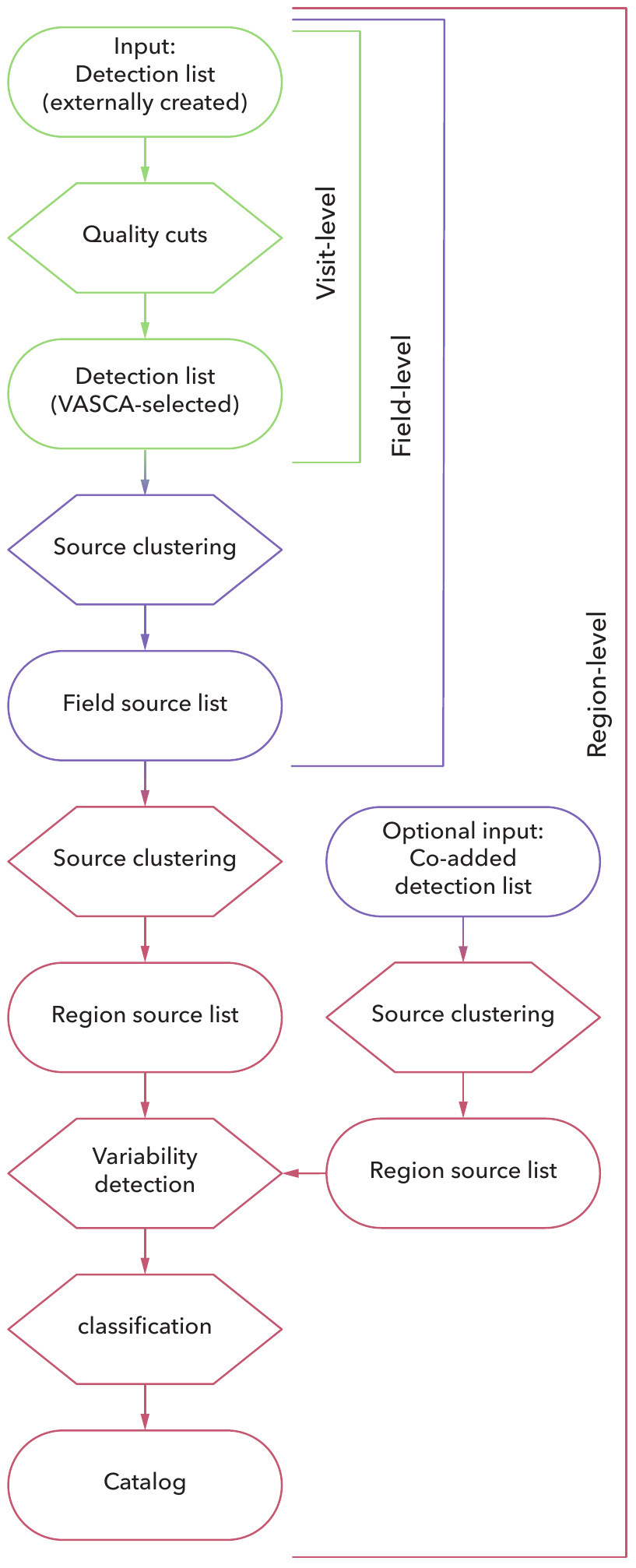}
\caption{Schematic of the VASCA processing flow.\label{fig:processing_flow_thin}}
\end{figure}

\clearpage
\FloatBarrier
\section{Catalog tables content} \label{sec:example_src}

\begin{table*}[h]
    \caption{
        Columns of the ``SOURCES'' table of the 1UVA catalog.
        \label{tab:cat}
    }
    \centering       
    \begin{tabular}{l l l }
        \hline\hline 
        \multicolumn{1}{c}{Name} & \multicolumn{1}{c}{Description} & \multicolumn{1}{c}{Unit} \\
        \hline
        SRC\_NAME & VASCA catalog source name &  \\
        NR\_DET & Number of detections &  \\
        RA & Sky coordinate Right Ascension (J2000) & degree \\
        DEC & Sky coordinate Declination (J2000) & degree \\
        POS\_ERR & Sky coordinate position error & arcsec \\
        POS\_XV & Sky position excess variance & arcsec$^2$ \\
        POS\_VAR & Sky position variance & arcsec$^2$ \\
        POS\_CPVAL & Sky position quality &  \\
        POS\_RCHIQ & Sky position reduced chisquared of the constant mean &  \\
        FLUX & Flux density & $\mu$Jy \\
        FLUX\_ERR & Flux density error & $\mu$Jy\\
        FLUX\_NXV & Flux normalized excess variance &  \\
        FLUX\_VAR & Flux variance & 10$^{-12}$ Jy$^2$ \\
        FLUX\_CPVAL & Probability value for a constant flux from the chisquare test &  \\
        FLUX\_RCHIQ & Flux reduced chisquared of the constant mean &  \\
        COADD\_SRC\_ID & Co-add source ID number &  \\
        COADD\_FFACTOR & Source flux divided by flux of the associated co-add source &  \\
        COADD\_FDIFF\_S2N & Signal to noise of the flux difference &  \\
        RG\_SRC\_ID & Region source ID number &  \\
        NR\_FD\_SRCS & Number of field sources &  \\
        HR & Flux hardness ratio, only simultaneous detections considered &  \\
        HR\_ERR & Flux hardness ratio error &  \\
        OGRP\_SIMBAD & SIMBAD source type group in VASCA &  \\
        OTYPE\_SIMBAD & SIMBAD source type &  \\
        MAIN\_ID\_SIMBAD & SIMBAD main ID &  \\
        SOURCE\_GAIADR3 & \textit{Gaia} DR3 source ID &  \\
        WDJNAME\_GAIAEDR3\_WD & \textit{Gaia}-EDR3-WD object name &  \\
        OBJID\_GFCAT & GFCAT object ID &  \\
        LS\_PEAK\_PVAL & LombScargle power probability value &  \\
        LS\_PEAK\_FREQ & LombScargle peak frequency & d$^{-1}$ \\
        \hline                  
    \end{tabular}
\end{table*}

\begin{table*}[h]
    \caption{
        Tables of the 1UVA catalog.
        \label{tab:tables}
    }
    \centering       
    \begin{tabular}{l l }
        \hline\hline
        \multicolumn{1}{c}{Name} & \multicolumn{1}{c}{Description} \\
        \hline
        SOURCES &  Properties of the 1UVA sources\\
        DETECTIONS & Properties of detections of 1UVA sources \\
        FIELDS & Properties of the analysed fields \\
        VISITS & Properties of the analysed visits \\
        FILTERS & Observation filter description \\
        COADD\_SOURCES &  Properties of the co-add sources \\
        \hline                  
    \end{tabular}
\end{table*}

%\clearpage
% tab:tables
%\input{tables/1uva_tables}
%\pagebreak
%\newpage % does nothing :(
% tab:cat
%\input{tables/1uva_columns}

%
%-----------------------------
\clearpage
\FloatBarrier
\section{Periodogram of White Dwarf J004917.14-252556.81}\label{sec:puls}

\begin{figure*}[h]
\centering
\includegraphics[width=0.95\linewidth]{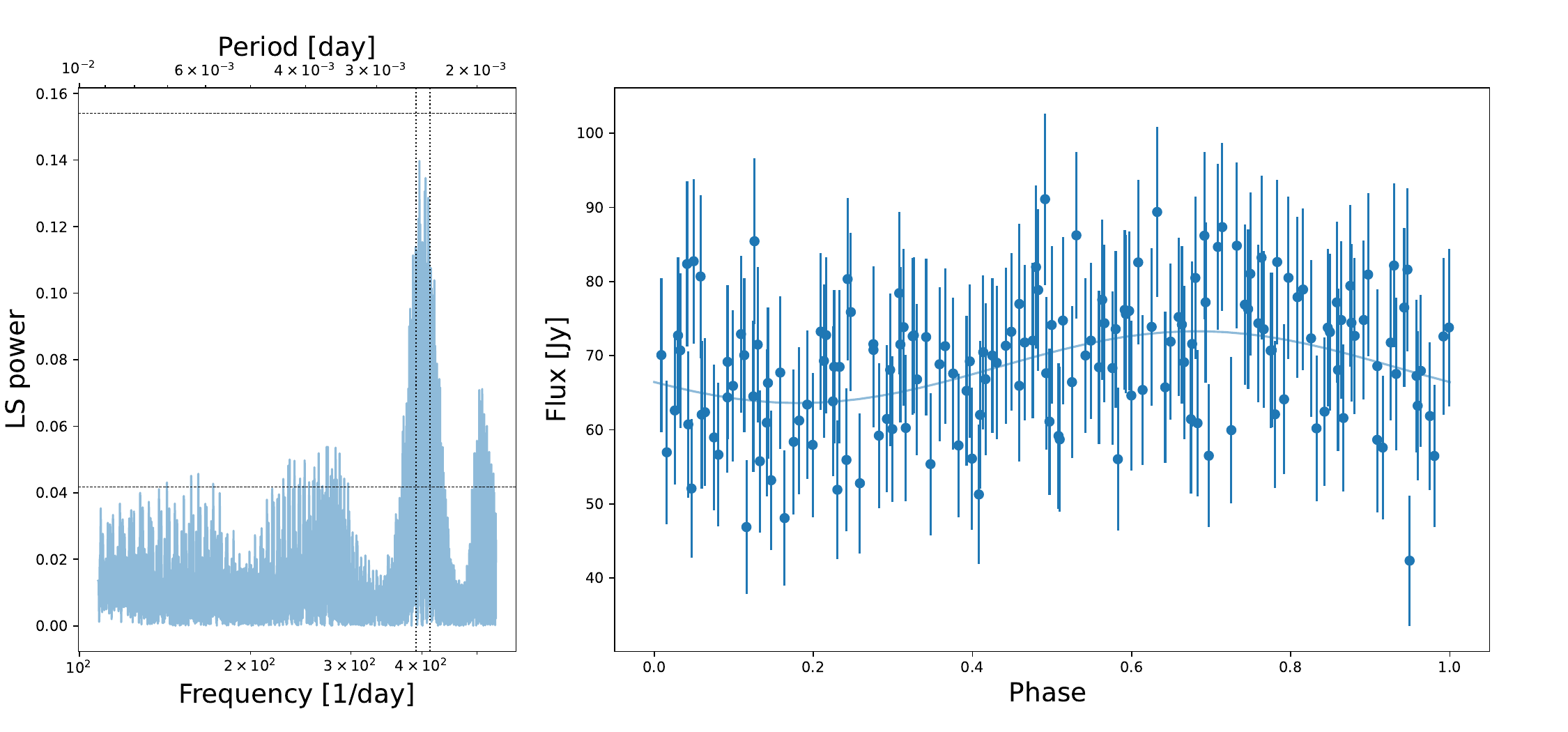}
\caption{ \textbf{Left:} Lomb Scargle periodogram of the source 1UVA~J04917.1-252556.5. It was calculated from a UV light curve in a 40 second time binning, restricted to the time range between MJD 55104.57558--55106.78567. Dashed horizontal lines mark the 2 and 3 $\sigma$ confidence level calculated following \citetads{2008MNRAS.385.1279B}. Vertical dotted lines mark the periods previously found for this source \citepads{2023MNRAS.522.2181K}. For more information see the main text. \textbf{Right:} The phased light curve with the best-fit model (straight line).  \label{fig:puls}}
\end{figure*}
    
\end{appendix}

\end{document}